\documentclass[10pt]{iopart}

\expandafter\let\csname equation*\endcsname\relax
\expandafter\let\csname endequation*\endcsname\relax

\usepackage{iopams}
\usepackage{amssymb}
\usepackage{bm}

\usepackage[version=4]{mhchem}

\usepackage{graphicx}
\usepackage{xcolor}
\usepackage{array}
\usepackage{booktabs}
\usepackage{multirow}
\usepackage{dcolumn}
\usepackage{wasysym}

\usepackage[nocompress]{cite}
\usepackage[colorlinks=true,linkcolor=red,citecolor=blue,urlcolor=blue]{hyperref}
\usepackage{cleveref}
\crefname{appendix}{}{}

\usepackage[protrusion=true,expansion=true,final]{microtype}
\usepackage[normalem]{ulem}

\newcommand{\mn}{\ce{Na2Mn3O7}}
\newcommand{\onlinecite}[1]{\cite{#1}}

\makeatletter
\long\def\@makecaption#1#2{%
  \vskip 8pt
  {\small\noindent\parbox{\linewidth}{\textbf{#1.} #2\parfillskip=0pt\relax\par}}%
  \vskip 10pt}
\makeatother

\makeatletter
\long\def\@makefntext#1{\parindent 1em\noindent
\makebox[1em][l]{\footnotesize\rm$\m@th{\arabic{footnote}}$}%
\footnotesize\rm \nobreakspace #1}
\def\@makefnmark{\textsuperscript{\hbox{${\arabic{footnote}}\m@th$}}}
\def\@thefnmark{\arabic{footnote}}
\makeatother

\begin{document}

\title[Crystallography-driven molecularization of a spin-$3/2$ magnet]{Crystallography-driven molecularization of a two-dimensional spin-$3/2$ magnet}

\author{Hari Borutta$^{1}$, Tobias M\"uller$^{2}$, Ronny Thomale$^{2,1}$, Harald O.\ Jeschke$^{3,1}$ and Yasir Iqbal$^{1}$}

\address{$^{1}$ Department of Physics, Indian Institute of Technology Madras, Chennai 600036, India}
\address{$^{2}$ Institut f\"ur Theoretische Physik und Astrophysik and W\"urzburg-Dresden Cluster of Excellence ct.qmat, Julius-Maximilians-Universit\"at W\"urzburg, Am Hubland, Campus S\"ud, W\"urzburg 97074, Germany}
\address{$^{3}$ Research Institute for Interdisciplinary Science, Okayama University, Okayama 700-8530, Japan}

\ead{yiqbal@physics.iitm.ac.in}

\begin{abstract}
Large-spin two-dimensional magnets are generally expected to develop conventional
long-range order once the dominant exchange scale becomes appreciable.
The layered spin-$3/2$ maple-leaf compound Na$_2$Mn$_3$O$_7$ defies this
expectation: despite sizable antiferromagnetic interactions and no evident
disorder, it exhibits no magnetic ordering and displays two well-separated
thermodynamic crossover scales.
We show that this behavior originates from a crystallography-driven
molecularization of the magnetic degrees of freedom.
The low-symmetry structure partitions the Mn sublattice into inequivalent
exchange pathways, generating a pronounced hierarchy that nearly isolates
antiferromagnetic hexagons.
Magnetic correlations therefore develop in two stages: first within individual
hexagons at a scale set by the dominant exchange, and only at much lower
temperatures do frustrated inter-hexagon couplings attempt to establish
coherence across the lattice.
While isolated hexagons reproduce the two-step thermodynamic structure, the
experimentally relevant temperature scales emerge only once the hexagons are
embedded in the frustrated two-dimensional network.
The resulting quantum ground state is magnetically disordered, characterized by
strong intra-hexagon correlations and rapidly decaying inter-hexagon
correlations.
These results identify crystallographic inequivalence as a materials-level
mechanism for stabilizing molecularized and quantum-disordered states even in
large-spin two-dimensional magnets.
\end{abstract}

\maketitle

\section*{Introduction}

Frustrated quantum magnets are often introduced through a simple expectation:
increase the spin length, and quantum fluctuations weaken, so conventional magnetic
order should become more robust~\cite{Balents2010,Richter2004}. Yet instructive
counterexamples—from dimerized quantum paramagnets to proximate spin-liquid
regimes—arise precisely when the lattice and exchange network reorganize the
low-energy degrees of freedom into local motifs (dimers, plaquettes, or larger
clusters) rather than individual spins. The paradigmatic
Shastry--Sutherland model~\cite{Shastry1981} illustrates how an exchange hierarchy can
stabilize an exact product of singlets and how small perturbations unlock competing
phases and quantum criticality~\cite{deconf-SSM,Jimenez2021,Shi2022,Liu-2024,Qian-2025}.
This broader lesson motivates a timely question for materials-based frustrated
magnetism: when a real compound refuses to order, does it realize a genuinely
two-dimensional frustrated magnet, or do crystallography and exchange hierarchy
repartition the degrees of freedom into emergent clusters whose thermodynamic
signatures survive even when long-range order is suppressed?

The maple-leaf lattice (MLL)~\cite{Betts-1995} provides a particularly sharp arena
in which to pose this question. As a one-seventh site-depleted triangular lattice
with coordination number five, it interpolates between triangular and kagome
geometries and supports diverse classical and quantum
phases~\cite{Ebert-2026,Gresista-2025,Schmoll-2025,Gembe-2024}. On the theory side,
the MLL hosts an exact singlet product ground state for minimal anisotropy,
placing it next to the Shastry--Sutherland model among exceptional uniform
tilings~\cite{Ghosh2022}. Beyond this point, modern approaches have mapped out
competing magnetic orders, magnetization plateaus, and candidate
quantum-disordered regimes~\cite{Ghosh2023,Ghosh2024,Nyckees-2025,Gresista-2025,
Beck2024,Hutak2025,Ebert-2026,Schmoll-2025,Gembe-2024}. Experimentally,
MLL realizations occur in a growing family of natural and synthetic compounds,
including bluebellite~\cite{Mills2014}, mojaveite~\cite{Mills2014},
fuettererite~\cite{Kampf2013}, sabelliite~\cite{Olmi1995}, spangolite
\cite{Penfield1890,Miers-1893,Frondel-1949,Hawthorne1993,Fennell2011,Aguilar-2025},
and semi-classical MLL antiferromagnets such as
MgMn$_3$O$_7\cdot$3H$_2$O~\cite{Haraguchi2018}. In real materials, however,
structural distortions and inequivalent exchange paths are the rule rather than
the exception~\cite{Inosov2018,Norman2018,Haraguchi2018,Haraguchi2021,
Ghosh-bluebellite-2024,Fennell2011,Mills2014,Schmoll-2025_span,Ghosh-2025}.
These developments sharpen a central puzzle: how robust are MLL phases to the
structural inequivalence and further-neighbor exchanges that inevitably occur
in real compounds?

A striking case is the layered manganate \mn~\cite{Venkatesh2020,Saha2023}.
While most attention has focused on spin-$1/2$ copper-based realizations
\cite{Fennell2011,Haraguchi2021}, \mn\ hosts spin-$3/2$ moments on
$\mathrm{Mn^{4+}}$ and would naively be expected to order once the dominant
exchange scale becomes appreciable. Instead, experiments report no long-range
magnetic order down to the lowest measured temperatures, accompanied by
pronounced short-range correlations developing below
$\sim100~\mathrm{K}$~\cite{Venkatesh2020,Saha2023}. Its thermodynamics exhibit
two clearly separated crossover scales: a broad maximum in the uniform
susceptibility near $110$--$120~\mathrm{K}$ and a lower-temperature enhancement
of the magnetic specific heat around $60$--$70~\mathrm{K}$. Such a separation
of scales is difficult to reconcile with a simple uniform lattice magnet and
instead points to strong correlations forming on intermediate length scales.

More broadly, Na$_2$Mn$_3$O$_7$ raises a conceptual question extending beyond
the MLL itself: can crystallographic inequivalence reorganize a nominally
two-dimensional magnet into an array of emergent magnetic “molecules,” whose
internal correlations dominate thermodynamics and suppress long-range order
even for large spins? In what follows, we show that this is precisely what
occurs. Starting from the experimentally resolved triclinic crystal structure
with three inequivalent manganese sites~\cite{Chang1985} [see Supplementary Note 1 for why Ref.~\onlinecite{Chang1985} is preferred over Refs.~\onlinecite{Raekelboom2001, Song2019}], we derive a
microscopic Heisenberg Hamiltonian via \emph{ab initio} energy
mapping~\cite{Jeschke2011,Jeschke2013,Ghosh2019}. The resulting in-plane model
is strongly nonuniform: three dominant antiferromagnetic exchanges nearly
isolate hexagonal plaquettes, while weaker and mostly ferromagnetic couplings
frustrate inter-hexagon coherence. This hierarchy naturally localizes strong
antiferromagnetic correlations on hexagons, reminiscent of other
exchange-hierarchical magnets~\cite{BOT-Sachdev,Normand2011,Ruegg2005,
Jimenez2021}. Magnetic correlations thus develop in two stages: first within hexagons at a scale set by the dominant intra-hexagon couplings, and only at much lower temperatures do frustrated inter-hexagon links attempt to establish lattice coherence.

We combine classical Monte Carlo, exact diagonalization of isolated
spin-$3/2$ hexagons, and pseudofermion functional renormalization group
calculations to determine the thermodynamic signatures and quantum fate of
this hierarchy. While isolated hexagons robustly generate two crossover
scales, the experimentally observed temperature positions emerge only once
the hexagons are embedded in the frustrated two-dimensional network.
Quantum fluctuations suppress the incipient classical ordering tendency and
stabilize a magnetically disordered ground state with pronounced intra-hexagon
correlations and rapidly decaying inter-hexagon correlations. The associated
momentum-resolved correlations are consistent with available
powder-averaged neutron-scattering profile~\cite{Saha2023}.

These results sharpen the broader message of current MLL research:
structural inequivalence does not merely perturb the ideal phase diagram; it
can qualitatively repartition the degrees of freedom and stabilize
quantum-disordered behavior even for comparatively large spins.
Our explanation relies on a translation-invariant Hamiltonian extracted from
the experimentally resolved crystal structure and does not require quenched
randomness~\cite{ShimokawaKawamura2015,UematsuKawamura2017}.
Nor is \mn\ simply a weakly coupled molecular magnet:
the thermodynamic maxima are strongly renormalized relative to isolated-hexagon
scales, and the momentum-space correlations reflect an extended frustrated
network rather than independent cluster form factors~\cite{GatteschiSessoliVillain2006,Waldmann2007}.
Crucially, the exchange hierarchy emerges directly from crystallographic
inequivalence and cannot be adiabatically deformed to a uniform
two-dimensional lattice.

In \mn{}, the ``molecularization'' of magnetism is \emph{crystallography-enforced}:
a translation-invariant, low-symmetry exchange network generates a robust hierarchy
that reorganizes the low-energy degrees of freedom into emergent hexagonal units.
Importantly, the experimentally relevant crossover scales are \emph{not} set by an
autonomous hexagon spectrum; they arise only once these units are embedded in the
frustrated two-dimensional network, which collectively renormalizes the thermodynamic
response to lower temperatures. This provides a materials-level route to cluster-dominated quantum paramagnets beyond
the spin-$1/2$ paradigm.

\begin{figure}[t]
    \includegraphics[width=\textwidth]{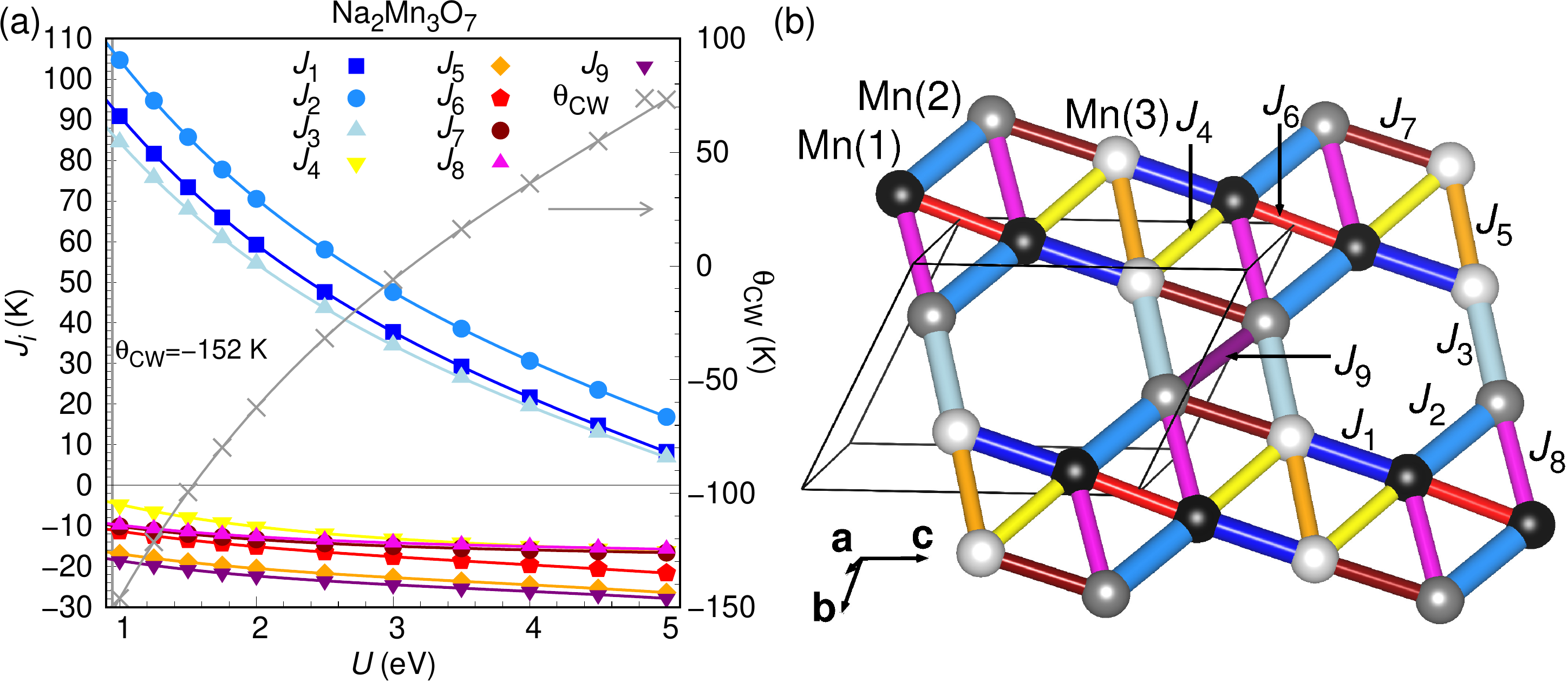}
    \caption{Hamiltonian of {\mn} determined by DFT based energy mapping. (a) Nine exchange interactions of {\mn} making up the maple leaf lattice as function of on-site interaction strength $U$, at fixed Hund's rule coupling $J_{\rm H}=0.76$\,eV. The vertical line indicates the value $U=0.94$\,eV at which the Heisenberg Hamiltonian parameters yield the Curie-Weiss temperature $T=-152$\,K determined by Venkatesh {\it et al.}~\cite{Venkatesh2020} from the experimental susceptibility. (b) Structure of the {\mn} lattice with the nine "nearest neighbor" exchange paths constituting the distorted maple leaf lattice. Strong antiferromagnetic couplings $J_1$--$J_3$ stabilize local N\'eel
order within hexagonal plaquettes. These hexagons are coupled by weaker, predominantly ferromagnetic interactions, which frustrate coherent inter-hexagon alignment and give rise to a stripe-like classical ordering tendency at an incommensurate wave vector which is classically determined to be $\mathbf{k}_0$ [see Eq.~\eqref{eqn:k_classical}]. The resulting classical state is only weakly stabilized, with the residual inter-hexagon coherence scale remaining small compared with the dominant intra-hexagon exchange scale.
}
    \label{fig:couplings}
\end{figure}

\begin{table}[htb]
\centering
\caption{
Exchange paths and Heisenberg Hamiltonian couplings as determined from DFT
based energy mapping. The paths are identified by Mn--Mn distance measured in the
structure from Ref.~\onlinecite{Chang1985}, and they are labeled according to increasing distance.
Note that interlayer couplings \(J_{22}\) to \(J_{24}\) and \(J_{31}\) to \(J_{36}\) are disregarded here. An exploratory larger-scale DFT energy-mapping calculation using a four-fold supercell
with 24 inequivalent Mn spins indicates that the leading interlayer exchanges are all
sub-Kelvin in magnitude, comparable to or smaller than \(J_{25}\), thereby confirming
the excellent two-dimensionality of Na\(_2\)Mn\(_3\)O\(_7\) on the energy scales considered here.
Throughout this work, ``nearest neighbors'' refer to the nine shortest Mn--Mn bonds per
Mn site identified in the experimental crystal structure and used in the DFT total-energy mapping.
}
\label{tab:couplings_transposed}
\begin{tabular}{c|c|c|c}
\hline
  & value (K) & role & $d_{\rm Mn-Mn}$ (\AA) \\
\hline\hline
 $J_1$  & 93.0788(3)  & NN ($h_1$) & 2.7887 \\
 $J_2$  & 107.089(2)  & NN ($h_2$) & 2.80028 \\
 $J_3$  & 86.5117(4)  & NN ($h_3$) & 2.805 \\
 $J_4$  & -4.27245(5) & NN ($t_1$) & 2.93047 \\
 $J_5$  & -16.5873(7) & NN ($d_1$) & 2.93479 \\
 $J_6$  & -10.9994(6) & NN ($d_2$) & 2.94347 \\
 $J_7$  & -9.84737(2) & NN ($t_2$) & 2.95048 \\
 $J_8$  & -9.55351(2) & NN ($t_3$) & 2.95498 \\
 $J_9$  & -18.2468(5) & NN ($d_3$) & 2.9606 \\
\hline
 $J_{10}$ & -7.24539(2) & 2NN($\varhexagon$) & 4.82937 \\
 $J_{11}$ & -8.00611(2) & 2NN($\varhexagon$) & 4.83238 \\
 $J_{12}$ & -4.98996(5) & 2NN($\varhexagon$) & 4.8764 \\
 $J_{13}$ & -14.1377(6) & 2NN & 4.9108 \\
 $J_{14}$ & -12.8032(7) & 2NN & 4.93777 \\
 $J_{15}$ & -11.8711(5) & 2NN & 4.94802 \\
 $J_{16}$ & -3.63742(2) & 2NN & 5.00436 \\
 $J_{17}$ & -2.52701(4) & 2NN & 5.02042 \\
 $J_{18}$ & -2.16424(2) & 2NN & 5.0209 \\
\hline
 $J_{19}$ & -8.32332(5) & 3NN($\varhexagon$) & 5.55231 \\
 $J_{20}$ & -8.76239(7) & 3NN($\varhexagon$) & 5.61725 \\
 $J_{21}$ & -9.4215(6)  & 3NN($\varhexagon$) & 5.61771 \\
 $J_{25}$ & 0.445469(4) & 3NN ($J_2{+}J_4$) & 5.73021 \\
 $J_{26}$ & 3.09456(6)  & 3NN ($J_1{+}J_6$) & 5.73072 \\
 $J_{27}$ & -- & 3NN ($J_1{+}J_7$) & 5.73843 \\
 $J_{28}$ & -- & 3NN ($J_3{+}J_5$) & 5.7395 \\
 $J_{29}$ & -- & 3NN ($J_3{+}J_8$) & 5.75864 \\
 $J_{30}$ & -- & 3NN ($J_2{+}J_9$) & 5.75961 \\
\hline
\end{tabular}
\end{table}

\section*{Results}

\subsection*{Model Hamiltonian}

The low-temperature triclinic crystal structure of
Na$_2$Mn$_3$O$_7$ hosts three symmetry-inequivalent
Mn$^{4+}$ sites arranged on a distorted maple-leaf lattice.
Starting from this experimentally resolved structure,
we derive a microscopic spin Hamiltonian using
Density Functional Theory (DFT)-based energy mapping [see Methods and Supplementary Note~1].

The resulting model is well described by an isotropic
Heisenberg Hamiltonian,
\begin{equation}
H = \sum_{i<j} J_{ij}\, \mathbf S_i \cdot \mathbf S_j \, ,
\end{equation}
where $\mathbf S_i$ are $S=3/2$ spins on Mn sites.

The exchange network is strongly nonuniform.
Three dominant antiferromagnetic couplings
$J_1$–$J_3$ form nearly isolated hexagonal plaquettes,
defining the primary energy scale
$J_h \sim 100~\mathrm{K}$. As we show below, this hierarchy produces a \emph{hexagon-first, network-later} regime: local correlations form at $T_h\sim J_h$, while the experimentally observed crossover temperatures are shifted downward by the frustrated inter-hexagon network.

All remaining interactions are substantially weaker
and include mostly ferromagnetic contributions.
This establishes a pronounced hierarchy
\begin{equation}
J_1, J_2, J_3 \gg J_{\mathrm{inter}} \, ,
\end{equation}
which partitions the lattice into strongly correlated
hexagonal units weakly coupled by frustrated
inter-hexagon exchanges.

The extracted parameters reproduce the experimental
Curie--Weiss temperature $\theta_{\mathrm{CW}}$
and are robust against moderate variations of the
interaction parameters [Supplementary Note~1].
A complete list of exchange constants is provided in
Table~\ref{tab:couplings_transposed}.

While there is, to the best of our knowledge, no direct experimental evidence for sizeable spin-exchange anisotropy in {\mn}---such as single-crystal susceptibility anisotropy, ESR constraints, or an inelastic-neutron signature requiring an anisotropic spin Hamiltonian---symmetry does not allow us to exclude Dzyaloshinskii--Moriya interactions or symmetric anisotropic exchange on all bonds. In particular, a DM vector is forbidden only when the corresponding bond has an inversion center at its midpoint; otherwise antisymmetric exchange is allowed in principle. The magnetic ion is \ce{Mn^{4+}}, i.e. a $3d^3$ ion in an octahedral crystal field, whose ground state is orbitally nondegenerate; the orbital moment is therefore largely quenched. This provides a physical basis for using an isotropic Heisenberg model as the minimal microscopic description. Our claim is not that anisotropies are symmetry-forbidden or exactly absent, but rather that the dominant exchange hierarchy and the associated hexagon-first physics already emerge at the isotropic level. Weak anisotropic exchanges, if present, are expected primarily to affect the lowest coherence scale, where several competing classical tendencies are close in energy, while the main physics in {\mn} remains controlled by the dominant hexagon-forming antiferromagnetic couplings.

\begin{figure}[htb]
    \includegraphics[width=\linewidth]{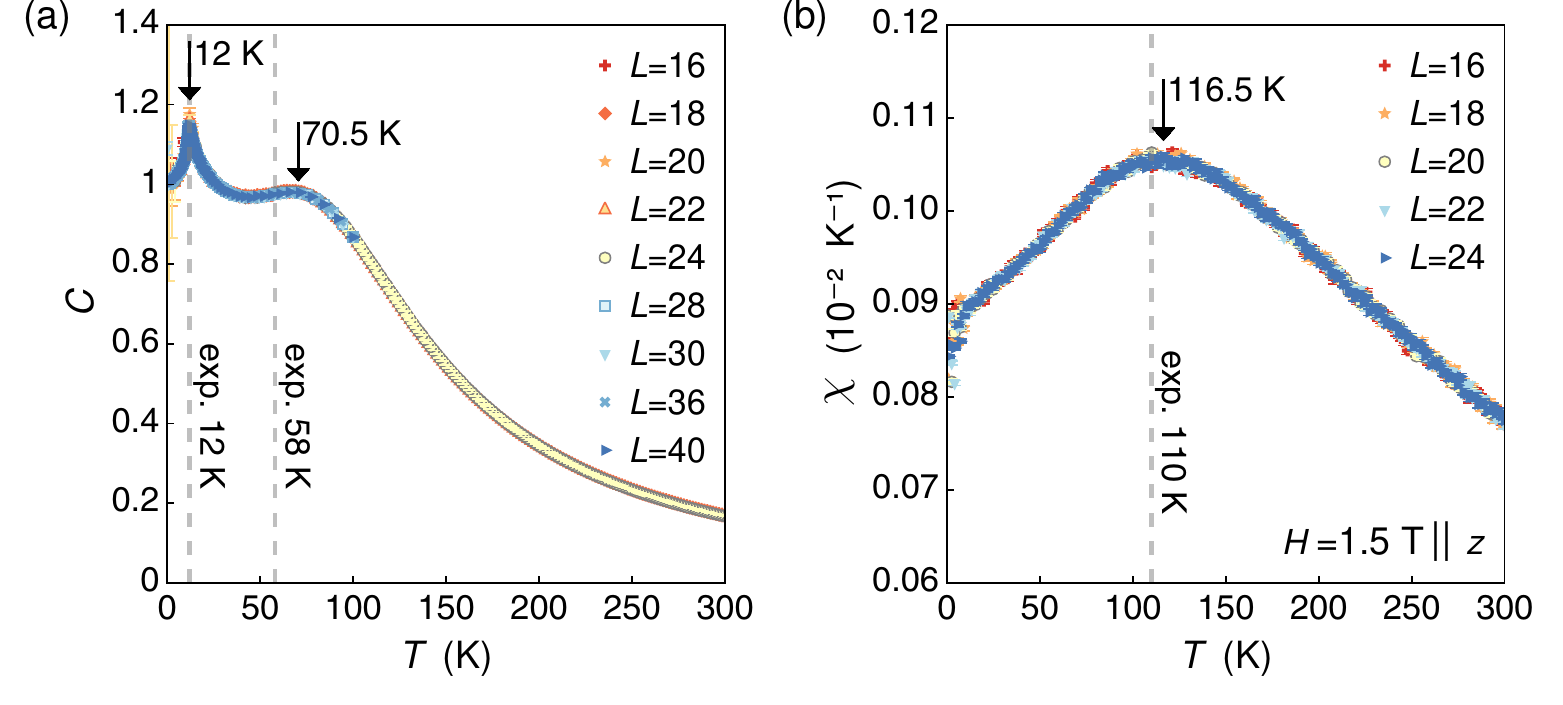}
    \caption{
Bulk thermodynamic properties obtained from cMC simulations of the
full Heisenberg model given in Table~\ref{tab:couplings_transposed}.
(a) Magnetic specific heat $C(T)$ for several system sizes of $L\times L$ unit cells with periodic
boundary conditions.
The specific heat increases upon cooling and develops a broad plateau around
$T\approx70$~K, followed by a weak low-temperature feature near $T\approx12$~K.
(b) dc magnetic susceptibility $\chi(T)$ calculated under an applied magnetic field
of $1.5$~T along the $z$ axis.
A broad maximum appears near $T\approx116$~K.
Dashed gray vertical markers indicate the temperatures of corresponding experimental
features reported in Ref.~\cite{Venkatesh2020}.
}
    \label{fig:CMC_C-chi_vs_T_U094}
\end{figure}

\subsection*{Luttinger--Tisza: incommensurate order with a shallow energy landscape}
As a classical reference, we analyze the full Heisenberg Hamiltonian using the extension of Luttinger--Tisza (LT) method, which allows us to treat lattices with multiple sublattices exactly~\cite{Schmidt2022}. Diagonalization of the Fourier-transformed dressed exchange matrix yields a unique minimum at the incommensurate wavevector
\begin{equation}
\label{eqn:k_classical}
\mathbf k_0 \simeq (0.50029,\,0.04252,\,0)\,\text{\AA}^{-1}
\end{equation}
with minimal eigenvalue
\begin{equation}
\lambda_{\min}(\mathbf{k}_0) = -206.415~\mathrm{K} \, ,
\end{equation}
corresponding to a classical energy per spin
\begin{equation}
E_{\mathrm{cl}} = -103.208~\mathrm{K} \, .
\end{equation}

The corresponding spin configuration is a single-$\mathbf{k}$ incommensurate coplanar spiral. For details see supplementary note 4.

Crucially, what is small here is \emph{not} the classical energy per spin but the
\emph{stiffness} of the ordering tendency: the LT spectrum has a shallow curvature near $\mathbf{k}_0$, implying a parametrically weak coherence scale compared to the dominant intra-hexagon exchange $J_h\sim100~\mathrm{K}$. The corresponding local LT dispersion is shown explicitly in Supplementary Fig.~S1, where the minimum appears as a shallow, anisotropic valley rather than a sharply confined basin.

\subsection*{Thermodynamics: hexagon scale vs network scale}

We next examine how this hierarchy manifests at finite temperature using classical Monte Carlo (cMC) simulations of the \emph{ab initio} Hamiltonian using all couplings specified in Table~\ref{tab:couplings_transposed}. This parameter set reproduces the experimental Curie--Weiss temperature and therefore serves as the baseline for comparison with thermodynamic measurements.

Figure~\ref{fig:CMC_C-chi_vs_T_U094}(a) shows the magnetic specific heat $C(T)$ for system sizes up to $40\times40$ unit cells.
Upon cooling, $C(T)$ develops a broad enhancement around $T \approx 70$~K, followed by a weak low-temperature feature near $12$~K.
Increasing system size does not sharpen this feature into a true anomaly.
Although the detailed shape differs from experiment, the characteristic temperature scales agree quantitatively with the broad enhancement and lower-temperature crossover reported in Ref.~\cite{Venkatesh2020}.
The finite low-temperature value of $C(T)$ reflects the classical equipartition contribution from quadratic fluctuations.
The weak $12$~K feature signals the incipient classical ordering tendency associated with the shallow LT landscape.

The magnetic susceptibility exhibits a broad maximum at $T \approx 116~\mathrm{K}$ [see Fig.~\ref{fig:CMC_C-chi_vs_T_U094}(b)], in quantitative agreement with experiment~\cite{Venkatesh2020,Saha2023}, confirming that the dominant exchange scale and the buildup of short-range correlations are captured already at the classical level.

\begin{figure}[t]
    \includegraphics[width=\linewidth]{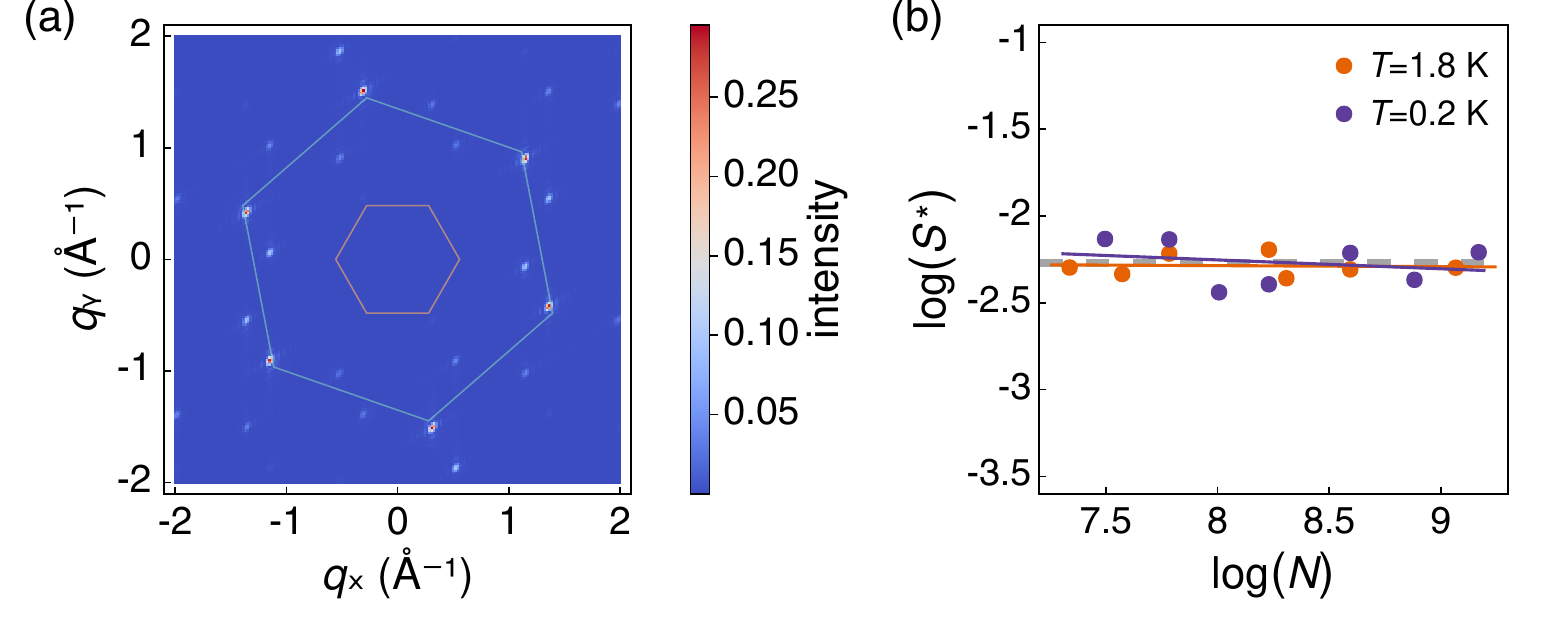}
    \caption{(a) Momentum-resolved equal-time spin structure factor $S(\mathbf{q})$ obtained from cMC simulations of the model given in Table~\ref{tab:couplings_transposed} at $T=1.8$~K, showing broad maxima at incommensurate wave vectors. Brillouin zone and extended Brillouin zone boundaries are shown by orange and green hexagons, respectively. The tilt of the extended Brillouin zone boundary with respect to the first Brillouin zone boundary seen in the figure originates from the geometry of the ideal flat $2$D maple-leaf lattice and from the chosen basis-vector pair used to define the corresponding extended zone. Put differently, the extended-zone construction is tied to the real-space orientation of the hexagon motif relative to the primitive lattice vectors. Even though a layer of the actual {\mn} material is distorted, the distortions are small enough that they do not qualitatively alter this geometric tilt.
(b) Finite-size scaling of the subdominant structure-factor peak associated with inter-hexagon coherence. We plot the peak intensity $S^\ast$ near $\mathbf{q}\simeq(1.1,-0.1)\,\mathrm{\AA}^{-1}$ as a function of the total number of sites $N=6L^2$ on a log-log scale, for $T=1.8$~K and $T=0.2$~K. The colored straight lines are linear fits to the corresponding data sets and are shown together with a zero-slope gray dashed line as a reference. The nearly flat behavior indicates that the peak intensity does not show the Bragg-like growth with system size expected for conventional long-range magnetic order. This supports the interpretation of the low-temperature cMC feature as a short-ranged incipient coherence scale rather than a finite-temperature ordering transition.
}
    \label{fig:sq-cmc_fin-size}
\end{figure}

The equal-time structure factor provides complementary information. Figure~\ref{fig:sq-cmc_fin-size}(a) shows $S(\mathbf{q})$ at $T=1.8$~K. Pronounced but broadened maxima appear near the corners of the extended Brillouin zone, with weaker subdominant features inside. Their finite width and weak size dependence indicate that, at the temperatures accessible in the simulations, the correlations remain short-ranged rather than developing into resolution-limited Bragg peaks. This is illustrated by the log-log finite-size analysis of the subdominant peak near $\mathbf{q}\simeq(1.1,-0.1)\,\mathrm{\AA}^{-1}$, shown in Fig.~\ref{fig:sq-cmc_fin-size}(b), for $T=1.8$~K and $T=0.2$~K. The peak intensity shows no proportional growth with system size, supporting the interpretation of the low-temperature feature as a soft incipient coherence scale rather than a finite-temperature transition. This is consistent with the Mermin--Wagner theorem, which forbids spontaneous breaking of continuous spin-rotation symmetry at finite temperature in a strictly two-dimensional isotropic Heisenberg model. The $T=0$ classical limit is nevertheless controlled by the LT minimum at $\mathbf{k}_0$, so the finite-temperature cMC data should be viewed as probing the approach toward a very soft classical ordering tendency rather than establishing a robust ordered phase.

\begin{figure}[t]
    \centering
    \includegraphics[width=\linewidth]{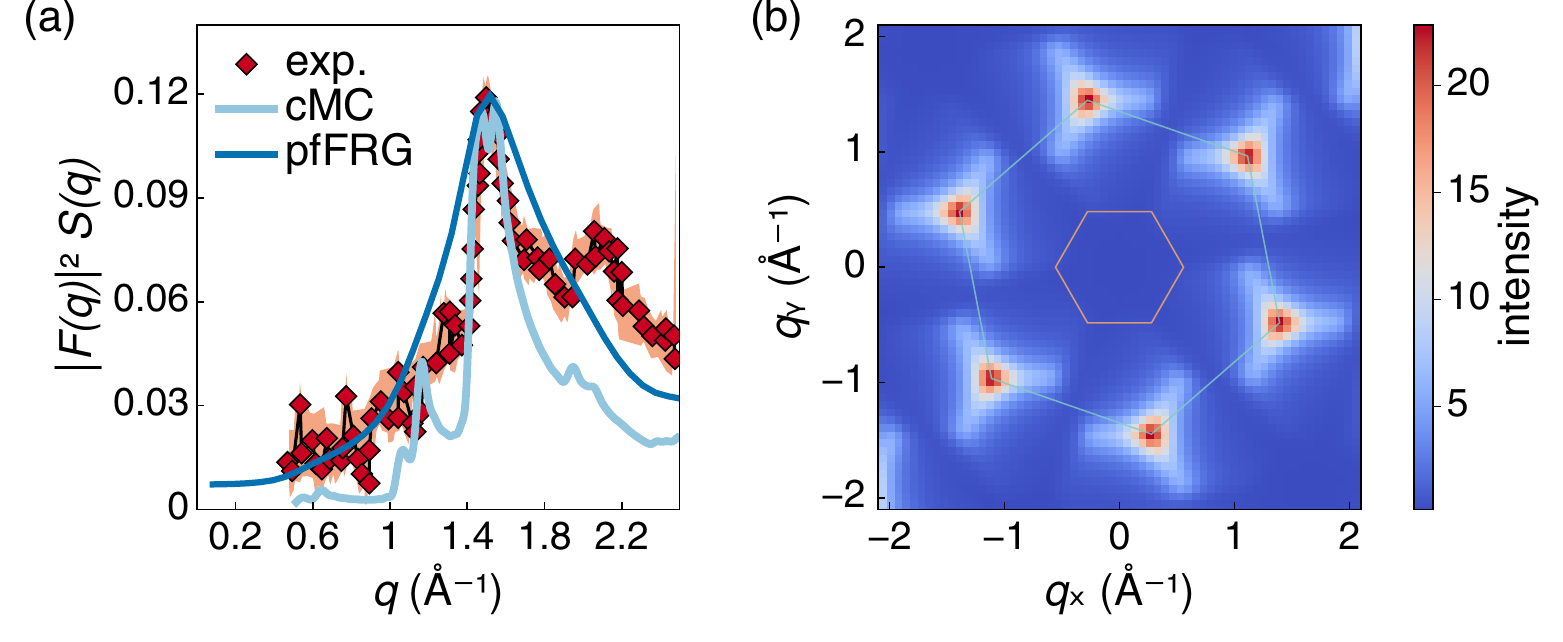}
    \caption{(a) Form-factor-modulated powder-averaged structure factor $|F(q)|^2 S(q),\; (q = |\mathbf{q}|)$, computed from cMC and pf-FRG at $T=1.8$~K. Experimental neutron-scattering data and associated error profile (orange shaded region) are taken from Ref.~\cite{Saha2023}. The broad features and absence of sharp Bragg peaks reflect dominant short-range correlations and suppressed long-range magnetic order. 
    (b) Static spin structure factor obtained from pf-FRG calculations. The structure factor displays broad maxima at wave vectors comparable to the classical results, yet remains smooth and finite, indicating that quantum fluctuations preserve the dominant short-range correlations while preventing their condensation into long-range order, consistent with a cluster-dominated quantum paramagnetic regime. The definitions of primitive lattice vectors and the basis vectors used to produce the above results are given in the Supplementary Note 3.}
    \label{fig:pwd_sq-pfFRG}
\end{figure}

The form-factor-modulated powder average [see Fig.~\ref{fig:pwd_sq-pfFRG}(a)] reproduces the dominant neutron-scattering peak position within experimental uncertainty~\cite{Saha2023}. This broad powder peak is dominated by the shorter-distance, stronger hexagon-scale correlations. The lower energy scale associated with frustrated inter-hexagon coherence does not generate a separate, sharply resolved powder feature in Fig.~\ref{fig:pwd_sq-pfFRG}(a). Rather, it enters more indirectly through the low-temperature redistribution and mild sharpening of diffuse spectral weight. This is not surprising, since powder averaging washes out directional information and the lower scale here reflects a soft incipient coherence tendency rather than a distinct second local motif.

A representative cMC snapshot on a finite lattice at $T=0.001$~K is shown in Supplementary Fig.~S2. At this very low temperature, thermal fluctuations are strongly reduced and the finite-size configuration displays a regular modulation compatible with the LT-selected incommensurate tendency. We regard this snapshot as an illustration of the fragile classical coherence tendency, not as evidence for a finite-temperature ordered phase. The absence of proportional growth in the finite-size scaling of the subdominant structure-factor peak, together with the lack of a sharp thermodynamic anomaly, supports the interpretation that the finite-temperature classical simulations remain in a short-ranged, incipient-coherence regime. Whether the physical $S=3/2$ quantum system condenses this fragile classical tendency requires an explicit quantum treatment, which is provided below by pf-FRG.

\subsection*{Cluster-dominated thermodynamics: hexagon-first physics}

The \emph{ab initio} exchange hierarchy provides a natural organizing principle for the thermodynamics.
The dominant antiferromagnetic couplings $J_1$--$J_3$ form strongly correlated hexagonal plaquettes, while all remaining exchanges are weaker and largely ferromagnetic.
Several weaker interactions act within individual hexagons ($J_{10}$--$J_{12}$ and $J_{19}$--$J_{21}$), reinforcing the separation between local and inter-hexagon scales.

To isolate the local building block, we solve the six-site $S=3/2$ hexagon by exact diagonalization. The spectrum exhibits a singlet ground state with energy $E_0=-277.18$~K per spin and a lowest triplet separated by $\Delta\sim10^2$~K. The resulting thermodynamics display two broad maxima: $T_{\max}^{\chi_{\varhexagon}}\simeq219$~K and $T_{\max}^{C_{\varhexagon}}\simeq148$~K.
Thus, the separation of thermodynamic scales is intrinsic to a single hexagon. Including intra-hexagon diagonal exchanges yields $E_0=-282$~K per spin and only weakly renormalized crossover scales ($T_{\max}^{\chi_{\varhexagon}}\simeq223$~K,
$T_{\max}^{C_{\varhexagon}}\simeq153$~K). All interactions confined to a hexagon therefore preserve the high-temperature scale separation. However, these isolated-cluster scales exceed those of the lattice response ($\sim116$~K in $\chi$, $\sim70$~K in $C$). The existence of two crossovers is a robust hexagon property, but their experimentally relevant positions require embedding in the frustrated two-dimensional network.

As an \emph{illustrative} (not quantitative) way to visualize how inter-hexagon couplings renormalize cluster thermodynamics, one may write an Random Phase Approximation (RPA)-type dressing,
\begin{equation}
\chi_{\mathrm{RPA}}(T)=\frac{\chi_{\varhexagon}(T)}{1-\lambda\,\chi_{\varhexagon}(T)} \, ,
\end{equation}
where $\lambda$ is the effective RPA parameter renormalizing the high-$T$ Curie–Weiss slope. Matching the high-temperature Curie--Weiss law $\theta_{\mathrm{CW}}\simeq -152$~K yields $\lambda\simeq403$~K.
This renormalizes the Curie--Weiss slope while preserving the intrinsic two-scale structure, illustrating that lattice couplings shift the thermodynamic scales without eliminating their hexagon origin, in agreement with findings from a recent numerical linked-cluster expansion study~\cite{Schafer-2026}. We stress that the quantitative scale renormalization is established by the full-lattice cMC results; the RPA expression is used only to emphasize the direction and mechanism of the shift.

The resulting picture is clear:
strong antiferromagnetic correlations form first on hexagons,
while frustrated inter-hexagon couplings suppress coherent order and shift thermodynamic anomalies to lower temperatures.
Unlike conventional molecular magnets, where clusters set the scale and the crystal embedding only fixes a much lower ordering temperature~\cite{GatteschiSessoliVillain2006,Waldmann2007,Schnack2018}, here the frustrated network is an active ingredient that renormalizes the thermodynamics downward. Na$_2$Mn$_3$O$_7$ thus realizes an emergent \emph{molecularization of a two-dimensional magnet}. Here we use ``molecularization'' operationally to denote a strong separation of scales: robust intra-hexagon correlations form at $T_h\sim J_h$, whereas inter-hexagon coherence is governed by a much smaller frustrated scale $T_{\mathrm{net}}\ll T_h$.

\begin{figure}[t]
    \includegraphics[width=\linewidth]{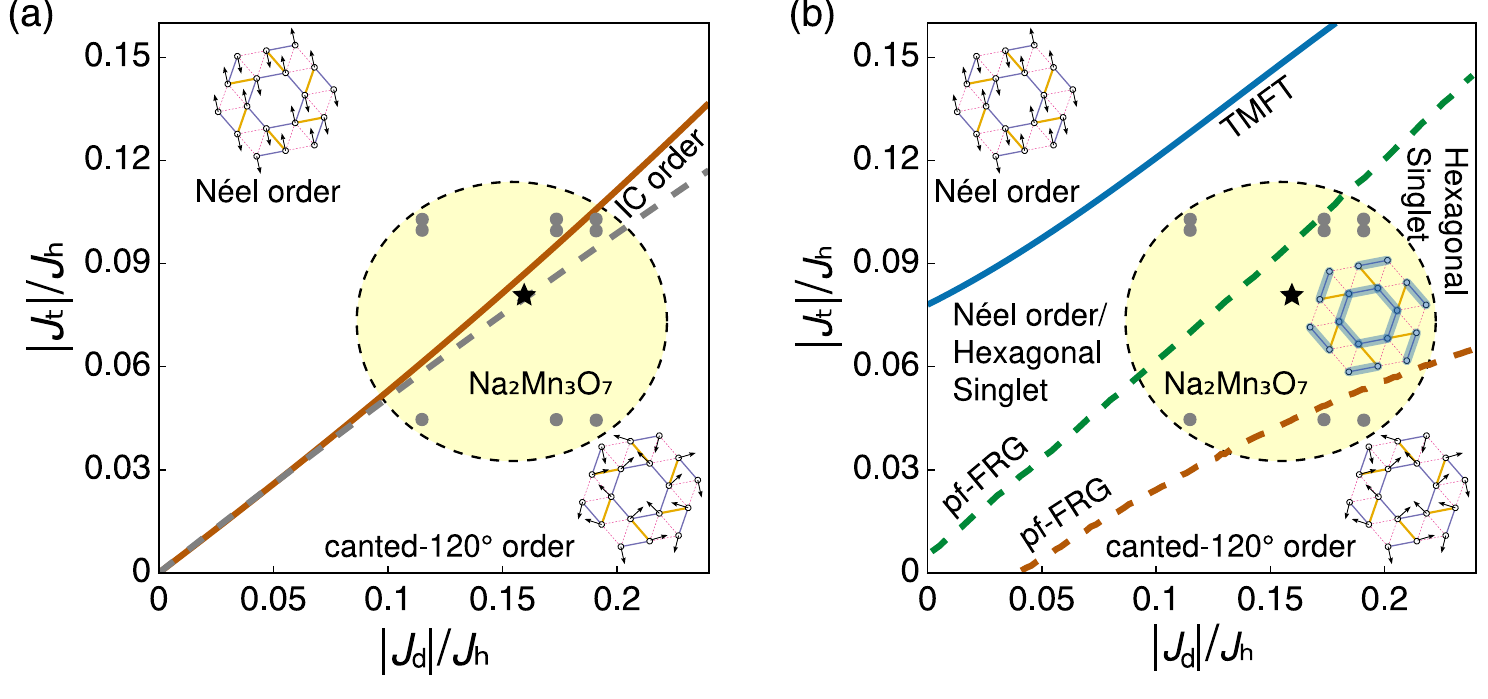}
\caption{\label{fig:phase_boundaries}
Phase boundaries of the symmetrized nearest-neighbor maple-leaf model and the location of Na$_2$Mn$_3$O$_7$ in coupling space. The point marked as star is for the averaged nearest-neighbor couplings $J_h, J_t, J_d$ obtained from Table~\ref{tab:couplings_transposed}, and the nine gray points are nine sets of nearest-neighbor $J_t-J_d$ combinations from the Table with the three nearest-neighbor $J_h$ averaged out. The region shaded in yellow is the best possible elliptical fit enclosing these ten coupling values.
(a) Classical phase diagram (Luttinger--Tisza / energy minimization) in terms of the averaged couplings $(J_h,J_t,J_d)$, showing competing N\'eel, canted-$120^\circ$, and incommensurate (IC) phases.
(b) Quantum phase boundaries from triplon mean-field theory (TMFT) [see Ref.~\onlinecite{Ghosh2024}] and pf-FRG. While the two approaches give different quantitative estimates for the extent of the magnetically disordered regime, both bracket a finite parameter window where dipolar order is suppressed for $S=1/2$; we use this window only as a qualitative upper bound on the corresponding regime expected for $S=3/2$; the Na$_2$Mn$_3$O$_7$ coupling ratios fall inside this bracketed region.}
    \label{fig:PB}
\end{figure}

Importantly, this is \emph{not} a weakly coupled molecular magnet.
Crystallography enforces a translation-invariant exchange hierarchy that places the lattice close to competing ordering tendencies, generating a parametrically small coherence scale and pushing the thermodynamic crossovers to temperatures far below the isolated-hexagon scales. It is worth noting that the two-step thermodynamic structures are not unique to extended lattices: finite antiferromagnetic clusters such as rings or polyhedral molecular nanomagnets also exhibit a susceptibility maximum at higher temperature than the dominant specific-heat maximum, reflecting discrete multiplet
physics \cite{GatteschiSessoliVillain2006,Schnack2018}. In Na$_2$Mn$_3$O$_7$, however, the short-range correlations are strongest on hexagons but extend beyond them, and their momentum-space fingerprints correspond to a correlated lattice rather
than independent molecular form factors, as addressed in the next section.

\begin{figure}[t]
\centering
    \includegraphics[width=0.75\linewidth]{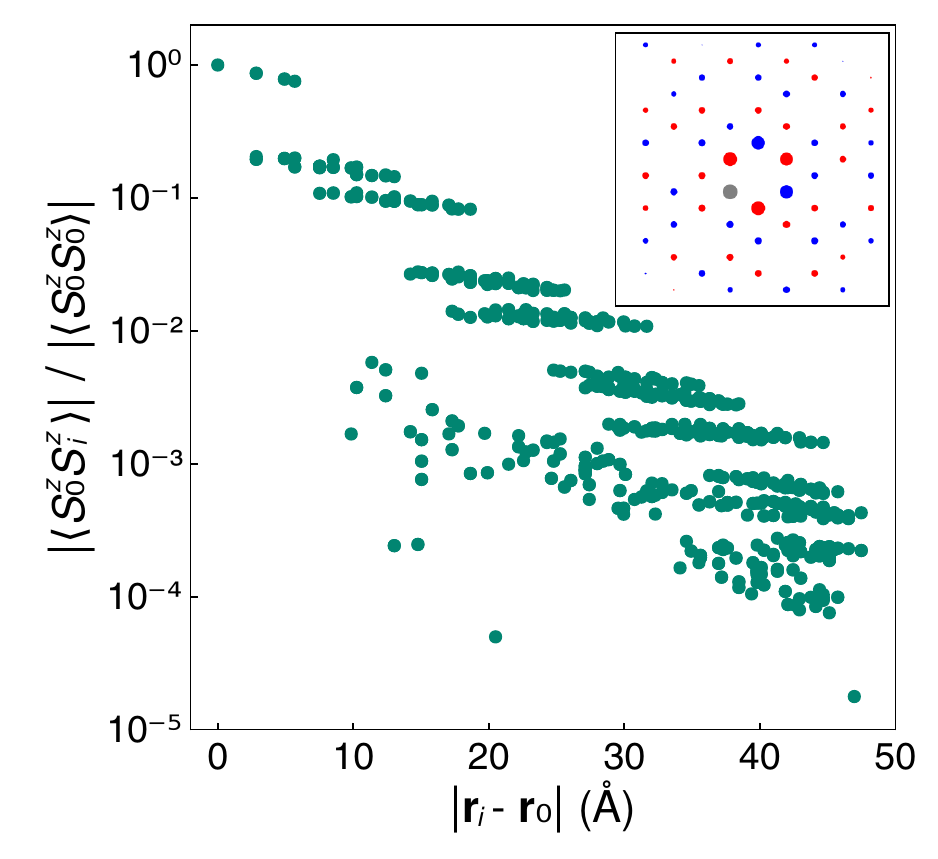}
\caption{Real-space correlations from the pf-FRG analysis (symmetrized nearest-neighbor model, supplemented by leading further-neighbor couplings).
Main panel: normalized equal-time correlations $|\langle S_0^z S_i^z\rangle|/|\langle S_0^z S_0^z\rangle|$ as a function of separation $|\mathbf{r}_i-\mathbf{r}_0|$, demonstrating strong antiferromagnetic correlations confined to individual hexagons and a rapid decay beyond the hexagonal unit.
Inset: map of $|\langle S_0^z S_i^z\rangle|$ on the lattice; the reference site $0$ is highlighted explicitly. The finite but rapidly decaying inter-hexagon correlations distinguish this regime from a lattice of independent magnetic molecules.
}
    \label{fig:real_space_spin_corr}
\end{figure}

\subsection*{Pseudofermion functional renormalization group: quantum melting of fragile coherence}

We now incorporate quantum fluctuations using pseudofermion functional renormalization group (pf-FRG) technique~\cite{Reuther-2010,Reuther-2011,Buessen-2019,
Kiese-2022,Muller-2024}.
To isolate the dominant competition while preserving the hierarchy, we symmetrize the nearest-neighbor couplings as
$J_h=(J_1+J_2+J_3)/3$,
$J_t=(J_4+J_7+J_8)/3$, and
$J_d=(J_5+J_6+J_9)/3$.
For the \emph{ab initio} parameters, $J_h\simeq95.6$~K, with modest anisotropy ($\delta_h\simeq0.21$, $\delta_{\mathrm{NN}}\simeq0.09$). Here, $\delta_h$ and $\delta_{\mathrm{NN}}$ are dimensionless anisotropy measures which quantify the deviation of the microscopic nearest-
neighbor model from the symmetrized description and are defined as
\begin{equation}
\delta_h =
\frac{\max(J_1,J_2,J_3)-\min(J_1,J_2,J_3)}{J_h}\,,
\end{equation}

\begin{equation}
\delta_{\mathrm{NN}} =
\frac{\sqrt{(J_1-J_h)^2+(J_2-J_h)^2+(J_3-J_h)^2}}{\sqrt{3}\,J_h}\,.
\end{equation}
and analogous expressions for the triangular and dimer bonds.
This preserves the dominant separation of scales between the different types of couplings, as described previously, retaining the hexagon-dominated physics.
Classically, this model lies near multiple phase boundaries [see Fig.~\ref{fig:PB}], reflecting strong frustration. This is still true, when accounting for deviations due to spatial anisotropy in the real material.
Including averaged further-neighbor couplings within and between hexagons, $J_{2\mathrm{NN}\varhexagon} = (J_{10}+J_{11}+J_{12})/3$ and $J_{3\mathrm{NN}\varhexagon} = (J_{19}+J_{20}+J_{21})/3$, respectively, as well as average second-nearest neighbor interaction $J_{2\mathrm{NN}} = (\sum_{i=13}^{18}  J_{i})/6$ connecting different hexagons, we compute the pf-FRG flow.
The maximum static susceptibility evolves smoothly over the entire cutoff range [see Supplementary Fig.~S3], with no divergence indicative of magnetic ordering.
This absence of flow breakdown persists with increasing spatial truncation, demonstrating suppression of dipolar long-range order.

The momentum-resolved structure factor [see Fig.~\ref{fig:pwd_sq-pfFRG}(b)] exhibits broad maxima at
$\mathbf q=\pm\mathbf k_0+\mathbf G$, coinciding with the cMC wavevector.
Quantum fluctuations broaden but do not shift these peaks, indicating that the ordering wavevector is fixed microscopically while coherence is melted~\cite{Schafer-2026}. Real-space correlations [see Fig.~\ref{fig:real_space_spin_corr}] are strongest within a hexagon and decay rapidly beyond it, yet remain finite over several lattice spacings~\cite{Gresista-2025}. This “strong-on-hexagons but not strictly confined” pattern demonstrates an interacting network of emergent clusters rather than independent molecular units. 

A direct experimental discriminator of this scenario is single-crystal diffuse neutron scattering: we predict broad intensity maxima at the symmetry-related wavevectors
$\pm\mathbf{k}_0+\mathbf{G}$ (with $\mathbf{G}$ reciprocal lattice vectors),
with substantial intrinsic width reflecting the finite correlation length and
without resolution-limited Bragg peaks.

The quantum phase boundaries for the pure $J_t$-$J_d$ model without further neighbor couplings have been investigated using triplon mean-field theory (TMFT) for both hexagonal singlet and dimerized ground states~\cite{Ghosh2024} and pf-FRG~\cite{Gresista-2025} for spin $1/2$. As shown in Fig.~\ref{fig:PB}, they bracket a finite disordered region, which we use as an upper bound for the region where we expect such physics for $S=3/2$.
The symmetrized Na$_2$Mn$_3$O$_7$ coupling ratios lie inside this window, even when taking into account deviations from this idealized limit in the real couplings, indicating that the quantum-disordered state arises from competition between nearby ordered phases rather than fine tuning. Quantum fluctuations thus melt the fragile classical coherence driven by weak inter-hexagon couplings. The resulting state is a \emph{cluster-dominated quantum paramagnet}: it is adiabatically connected to the limit of decoupled hexagons with singlet ground states, yet in the lattice it exhibits finite but short-ranged inter-hexagon correlations without dipolar long-range order. This provides the zero-temperature counterpart of the cluster-dominated thermodynamics established above.
While this conclusion is based on a symmetrized model, due to the hierarchical structure of the unsymmetrized couplings, we still expect it to hold qualitatively. The quantitative extent of the disordered region in parameter space, however, would be subject to the full anisotropy of the model.

\section*{Discussion}
\label{sec:conclusions}

Na$_2$Mn$_3$O$_7$ demonstrates that crystallographic inequivalence can act as a primary organizing principle for magnetism rather than a mere perturbation of ideal lattice models. The low-symmetry structure enforces a pronounced exchange hierarchy that repartitions the low-energy degrees of freedom from individual spins into strongly correlated hexagonal units. In doing so, it stabilizes quantum-disordered behavior even for comparatively large spins and identifies crystallography itself as a practical control knob for engineering cluster-based quantum paramagnets beyond the spin-$1/2$ paradigm.

We have established a coherent microscopic picture for the absence of long-range magnetic order in this layered spin-$3/2$ maple-leaf compound. Three dominant antiferromagnetic exchanges generate robust intra-hexagon correlations at a scale $T_h\sim J_h$, while substantially weaker, partly ferromagnetic couplings frustrate coherent inter-hexagon order at a parametrically smaller scale $T_{\mathrm{net}}$. This hierarchy naturally accounts for the experimentally observed two-step thermodynamic structure: a broad susceptibility maximum near $110$--$120$~K and a lower-temperature specific-heat enhancement around $60$--$70$~K.

Exact diagonalization establishes that the \emph{existence} of two thermodynamic crossovers is an intrinsic property of a single $S=3/2$ hexagon. The \emph{absolute} crossover temperatures, however, are not set by an autonomous cluster spectrum:
even after including all intra-hexagon couplings, the isolated-hexagon maxima remain far above experiment. The experimentally observed scales therefore require the frustrated embedding of hexagons into the two-dimensional network, which collectively renormalizes the response to lower temperatures. Na$_2$Mn$_3$O$_7$ is thus a \emph{molecularized two-dimensional magnet}---cluster physics dominates over a wide window, yet the lattice remains an active ingredient rather than a passive host.

Classically, the shallow energy landscape generated by weak inter-hexagon couplings leads to an incipient ordering tendency. Quantum fluctuations, treated using pf-FRG, suppress this fragile coherence and stabilize a magnetically disordered, cluster-dominated paramagnetic regime characterized by strong intra-hexagon correlations and rapidly decaying inter-hexagon correlations. The momentum-resolved correlations retain the classical ordering wavevector but remain broadened, indicating that frustration melts long-range order without altering the underlying exchange-imposed structure. From the perspective of the symmetrized nearest-neighbor model, the extracted coupling ratios place Na$_2$Mn$_3$O$_7$ near multiple competing ordered phases, showing that the disordered ground state arises from finely balanced competition rather than trivially weak interactions.

We do not intend this terminology to introduce a distinct new phase label; rather, it denotes a hierarchy-controlled quantum paramagnet continuously connected to the decoupled-hexagon singlet limit, while retaining measurable short-ranged inter-hexagon correlations.

Our cluster-based scenario suggests several direct experimental tests. Pressure or uniaxial strain should predominantly affect the weak inter-hexagon couplings and hence strongly modify the lowest-temperature crossover scale while leaving the high-temperature susceptibility maximum largely unchanged. Single-crystal diffuse neutron scattering would provide a sharper probe of the predicted incommensurate short-range correlations. Local probes such as $\mu$SR and NMR should detect persistent dynamics without static internal fields, governed primarily by intra-hexagon modes. Field-dependent thermodynamics may further reveal signatures of cluster multiplets broadened by frustrated inter-hexagon couplings.

For a more detailed theoretical understanding, the next step would be to derive an effective low-energy theory by projecting the full Hamiltonian onto the low-lying multiplet manifold of each hexagon. Such a theory would have to retain at least the hexagon singlet and the lowest magnetic excitations, and then derive the effective inter-hexagon couplings generated within this reduced Hilbert space. The difficulty is that the inter-hexagon couplings are weak enough to preserve the cluster hierarchy but not asymptotically negligible; moreover, frustration can generate multi-hexagon terms beyond a simple RPA dressing of the isolated-hexagon susceptibility. We therefore regard the RPA expression used above only as an illustrative indication of the downward renormalization of thermodynamic scales, not as a quantitative effective theory.

More broadly, this work sharpens an emerging lesson for frustrated magnets on nonuniform lattices: structural inequivalence and exchange hierarchy can qualitatively reorganize low-energy physics into emergent clusters and stabilize quantum-disordered behavior even for sizable spin. In Na$_2$Mn$_3$O$_7$, this mechanism suppresses long-range order despite substantial antiferromagnetic exchange and positions the system near competing ordered states. External control parameters such as pressure, strain, or selective chemical substitution—targeting the weak inter-hexagon couplings—offer promising routes toward magnetic order, quantum criticality, or novel cluster-based phases. More generally, low crystallographic symmetry, often regarded as a complication, can instead serve as a deliberate design principle for engineering exchange hierarchy and emergent cluster physics in two-dimensional magnets.

\section*{Methods}

\subsection*{Density functional theory and exchange extraction}
All DFT calculations were performed with the all-electron
full-potential local-orbital code FPLO~\cite{Koepernik1999} using the generalized
gradient approximation (GGA)~\cite{Perdew1996}. Strong electronic correlations on the
Mn $3d$ shell were treated within DFT+$U$~\cite{Liechtenstein1995} by varying the on-site
Coulomb interaction $U$, while keeping the Hund’s coupling fixed at
$J_{\rm H}=0.76$~eV~\cite{Mizokawa1996}. Exchange parameters were obtained from
DFT total-energy mapping [see Table~\ref{tab:couplings_transposed}].
The value $U=0.94$~eV was selected because it yields a Curie--Weiss temperature
consistent with experiment, $\theta_{\rm CW}=-152$~K~\cite{Venkatesh2020}.
All subsequent finite-temperature and quantum-many-body calculations reported in the
main text use this parameter set.

\subsection*{Classical Monte Carlo}
For the cMC simulations, we considered $L\times L$ unit-cell periodic lattices
with $L=16,18,20,22,24,28,30,36,40$.
Updates employed standard single-spin Metropolis--Hastings rotations.
A Monte Carlo sweep corresponds to $N$ attempted single-spin updates, with
$N=L^2$ the number of spins.

Thermodynamic quantities ($C$ and $\chi$) were computed on a uniform grid of
$190$ temperatures spanning $T=1$--$300$~K. At each temperature, measurements were
performed after thermalization using $15\times10^{6}$ sweeps for $300$--$50$~K,
$25\times10^{6}$ sweeps for $50$--$8$~K, and $10^{8}$ sweeps for $8$--$1$~K.
To access low-temperature quantities ($T\le 1.8$~K), including equal-time correlations,
structure factors, and representative low-$T$ configurations, we supplemented the
Metropolis updates by over-relaxation: after each Metropolis update at site $i$,
the spin was rotated by $180^\circ$ about the local effective field axis generated by
its neighbors. This energy-conserving move strongly reduces autocorrelation times in
the frustrated regime. For $16\times16$, the maximum autocorrelation time in the
equal-time structure-factor data is $\sim 500$ sweeps after including over-relaxation,
and we therefore used a binning size of $2048$ sweeps for reliable error estimates.
Since $S(\mathbf q)$ requires $O(N^2)$ operations per sweep, we measured it only once
every $\sim 2000$ sweeps and averaged over $1000$ effectively uncorrelated samples.

We evaluated the magnetic specific heat as
\begin{equation}
C(T)=\frac{1}{N}\frac{\langle E^{2}\rangle-\langle E\rangle^{2}}{T^{2}} \, ,
\end{equation}
and the dc susceptibility as
\begin{equation}
\chi=\frac{N}{T}\left(\langle M_z^{2}\rangle-\langle M_z\rangle^{2}\right) \, ,
\qquad
M_z=\frac{1}{N}\sum_i S_i^{z} \, ,
\end{equation}
with an applied field fixed to $1.5$~T. The equal-time structure factor was computed as
\begin{equation}
S(\mathbf{q})=
\frac{1}{N}\sum_{i,j}e^{i\mathbf{q}\cdot(\mathbf{r}_i-\mathbf{r}_j)}
\langle \mathbf{S}_i\cdot\mathbf{S}_j\rangle \, ,
\end{equation}
using the Mn positions $\mathbf r_i$ of the lattice convention adopted in the paper [see Supplementary Note 3],
and we further computed form-factor-modulated powder averages $|F(q)|^2 S(q)$, with
\begin{equation}
    \begin{split}
        F(q) =&\; 0.3760 e^{-12.5661\big(\frac{q}{4 \pi}\big)^2}
        + 0.6602 e^{-5.1329 \big( \frac{q}{4 \pi} \big)^2} \\
        &- 0.0372 e^{-0.5630 \big( \frac{q}{4 \pi} \big)^2}
        + 0.0011\,,
    \end{split}
\end{equation}
for Manganese ions~\cite{Brown2006}, for direct comparison to
experiment~\cite{Venkatesh2020,Saha2023}.

\subsection*{Pseudofermion functional renormalization group}
Quantum correlations and ordering tendencies were analyzed using pf-FRG in the \texttt{PFFRGSolver.jl} implementation~\cite{pffrgsolver}. Spins were represented by Abrikosov pseudofermions and the
renormalization-group flow was formulated at $T=0$ using a continuous infrared cutoff
$\Lambda$ in the fermionic propagator. We truncated the one-particle-irreducible vertex
hierarchy at the two-particle level and used the Katanin truncation scheme~\cite{Katanin-2004}
to feed back self-energy effects into the vertex flow~\cite{Reuther-2011,Buessen-2019}.

Spatial correlations were restricted to a finite real-space range by neglecting vertex
contributions beyond a maximal bond distance $L$, with values up to $L=18$ used to
assess convergence. The three-frequency dependence of the two-particle vertex was
discretized on an adaptive frequency grid of size $35\times40\times40$ (bosonic and two
fermionic frequency arguments, respectively). With these parameters, the pf-FRG flow
involves on the order of $9\times10^{7}$ coupled differential equations, solved
numerically using high-performance computing resources [see Ref.~\cite{Kiese-2022} for
implementation details].

The central observable is the static susceptibility
\begin{equation}
\chi_{ij}^{\Lambda,\mu\nu}=
\int_{0}^{\infty}d\tau\,
\langle \hat T_\tau \hat S_i^\mu(\tau)\hat S_j^\nu(0)\rangle^\Lambda \, ,
\end{equation}
and its momentum-space form
\begin{equation}
\chi^{\Lambda,zz}(\mathbf{k})=
\frac{1}{N}\sum_{ij}e^{i\mathbf{k}\cdot(\mathbf{r}_i-\mathbf{r}_j)}\chi_{ij}^{\Lambda,zz} \, .
\end{equation}
For SU(2)-symmetric Heisenberg models we restrict to $\chi^{zz}$ (all diagonal components
are equal and off-diagonal components vanish). The flow was integrated down to
$\Lambda/|J|=0.05$, with $|J|=\sqrt{\sum_i J_i^2}$ the overall exchange scale.
Dipolar magnetic long-range order is identified by a susceptibility divergence (“flow
breakdown”) at finite $\Lambda$~\cite{Reuther-2010,Buessen-2019}. To avoid false positives
from finite truncation, we used a conservative criterion based on the evolution of the
second derivative of the maximal susceptibility: a breakdown is diagnosed when
non-monotonic features emerge and become increasingly pronounced with increasing $L$,
following established practice~\cite{Gresista-2025_pyro,Gresista-2025,Gresista-2023}.
We emphasize that absence of a breakdown excludes dipolar order but may still allow
non-dipolar symmetry breaking (e.g.\ valence-bond or spin-nematic order), which would
require four-spin correlators beyond the present truncation.

\section*{Acknowledgements}
We thank Pratyay Ghosh for helpful discussions and for sharing the data from the triplon mean-field theory analysis of Ref.~\cite{Ghosh2024}. R.T. and T.M. acknowledge financial support by the Deutsche Forschungsgemeinschaft through ProjectID 258499086 – SFB 1170, through the Würzburg- Dresden Cluster of Excellence on Complexity and Topology in Quantum Matter – ctd.qmat Project-ID 390858490 – EXC 2147, and through the Research Unit QUAST, Project-ID 449872909 – FOR5249. H.O.J. acknowledges support through JSPS KAKENHI Grants No. 24H01668 and No. 25K08460. Part of the computation in this work has been done using the facilities of the Supercomputer Center, the Institute for Solid State Physics, the University of Tokyo. The work by Y.I. was performed in part at the Aspen Center for Physics, which is supported by a grant from the Simons Foundation (1161654, Troyer). This research was also supported in part by grant NSF PHY-2309135 to the Kavli Institute for Theoretical Physics and by the International Centre for Theoretical Sciences (ICTS) for participating in the Discussion Meeting - Fractionalized Quantum Matter (code: ICTS/DMFQM2025/07). Y.I. acknowledges support from the Abdus Salam International Centre for Theoretical Physics through the Associates Programme, from the Simons Foundation through Grant No.~284558FY19, from IIT Madras through the Institute of Eminence program for establishing QuCenDiEM (Project No. SP22231244CPETWOQCDHOC). H.O.J. and R.T. thank IIT Madras for a Visiting Faculty Fellow position under the IoE program.

\section*{Data Availability Statement} The data that support the findings of this study are openly available in Zenodo at \href{https://doi.org/10.5281/zenodo.20326372}{https://doi.org/10.5281/zenodo.20326372}.

\clearpage
\bibliography{mapleleaf}
\clearpage
\newcommand{\addpage}[1] {
\begin{figure*}
  \includegraphics[width=8.5in,page=#1]{SupplementaryMaterial.pdf}
\end{figure*}
}
\addtolength{\oddsidemargin}{-0.75in}
\addtolength{\evensidemargin}{-0.75in}
\addtolength{\topmargin}{-0.725in}
\addpage{1}
\addpage{2}
\addpage{3}
\addpage{4}
\addpage{5}
\addpage{6}
\addpage{7}
\addpage{8}
\addpage{9}
\addpage{10}

\end{document}